\documentclass[pra,tightenlines,twocolumn]{revtex4}

\usepackage{graphicx}
\usepackage{bbm}
\usepackage{amsfonts}
\usepackage{theorem}
\usepackage{amssymb}
\usepackage{amsmath}
\newcommand{\W}{{\mathcal W}}

\begin{document}



\title{The geometry of bipartite qutrits including bound entanglement}

\author{B. Baumgartner, B.C. Hiesmayr\footnote{Beatrix.Hiesmayr@univie.ac.at}, H. Narnhofer}
\address{Faculty of Physics, University of Vienna, Boltzmanngasse 5, 1090 Vienna, Austria}

\begin{abstract}
We investigate the state space of bipartite qutrits. We construct an
analog to the ``magic'' tetrahedron for bipartite qubits---a magic
simplex $\W$. It is formed by all convex combination of nine Bell
states which are constructed using Weyl operators. Due to the high
symmetry it is enough to consider certain typical slices through
$\W$. Via optimal entanglement witnesses we find regions of bound
entangled states.
\end{abstract}


 \maketitle

\section{Introduction}

Quantum entanglement is without doubt one of the most remarkable
features of quantum mechanics. It is the source of applications like
quantum cryptography, teleportation, dense coding or a possible
quantum computer. Up to now one does not yet have a good computable
way to distinguish entangled states from separable ones for higher
dimension or more particles and not much is known about the
structure of the set of separable states. In this letter we explore
a subset of bipartite qutrit states for which we find the
geometrical structure, surprisingly including a whole region of
bound entangled states.

To explore the geometrical structure, as is the main topic of many
works (e.g. Ref.~\cite{N06,VW00,Ovrum,PR04,W04,B04,VW99}), is of
great help in understanding the quantum features, develop quantum
measures and algorithms and hence to find future applications.

Since the seminal letter on distillation \cite{Bennett}, it was a
common expectation that all entangled bipartite states are
distillable, but already for bipartite qutrits one finds states
which are positive under partial transpose $PT$, i.e. have only
positive eigenvalues, called $PPT$ states, but cannot be distilled
\cite{Horodecki1998}, i.e by no local operation and classical
communication (LOCC) Alice and Bob can purify or distill this
bipartite mixed state into a maximally entangled one. These states
are called \textit{bound} entangled. Though a lot of examples are
found, e.g. Refs.~\cite{Augusiak, Breuer, Derkacz, Ruskai, Planat},
and even for thermal states (e.g. Ref.~\cite{Acin} and references
therein), general recipes to construct such kind of states are
lacking. The reason for the existence of these kind of states is
also unknown and still mysterious \cite{HorodeckiSummary}.

We start by considering a subset of bipartite qubits for which the
state space can be visualized via the ``magic'' tetrahedron
\cite{Horodecki,BNT02}. We find for bipartite qutrits a subset where
an analogous simplex can be drawn. We analyze the set of separable
states within and point out the similarities to bipartite qubits,
and discuss the differences, e.g. the polytope structure and bound
entanglement.

\section{Bipartite Qubits}

A single qubit state $\omega$ lives in a two dimensional Hilbert
space, i.e. ${\cal H}\equiv\mathbb{C}^2$, and any state can be
decomposed into the well known Pauli matrices
\begin{equation*} \label{qubitpauli}
\omega \;=\; \frac{1}{2} \left( \mathbbm{1}_2 + n_i\, \sigma^i
\right)
\end{equation*}
with the Bloch vector components $\vec{n} \in \mathbbm{R}^3$ and
$\sum_{i=1}^3 n_i^2 = \left| \vec{n} \right|^2 \leq 1$. For
$\left|\vec{n}\right|^2 < 1$ the state is mixed (corresponding to
Tr$\,\omega^2 < 1$) whereas for $\left|\vec{n}\right|^2 = 1$ the
state is pure (Tr$\,\omega^2 = 1$).

\begin{figure}
\center{\includegraphics[height=150pt,
keepaspectratio=true]{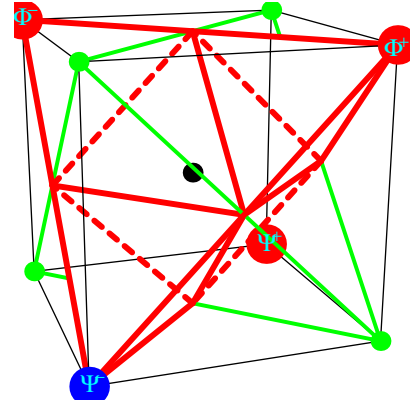}} \caption{(Color online) For two
qubits four orthogonal Bell states, $\psi^\pm,\phi^\pm$ can be used
to decompose every locally maximally mixed state and a geometric
picture can be drawn. The positivity condition forms a tetrahedron
(red) with the four Bell states at the corners of the cube and the
totally mixed state, the trace state, in the origin (black dot in
the middle). Via reflection $\vec{c}\rightarrow -\vec{c}$ one
obtains another tetrahedron (green) with reflected Bell states
located in the remaining corners of the cube. The intersection of
both tetrahedra gives an octahedron where all points inside and at
the
surface represent separable states. 
}\label{qubitstetrahedron}
\end{figure}

The density matrix of $2$--qubits $\rho$ on ${\mathbb C}^2 \otimes
{\mathbb C}^2$ is usually obtained by calculating its elements in
the standard product basis, i.e.
$|00\rangle,|01\rangle,|10\rangle,|11\rangle$. Alternatively, we can
write any $2$--qubit density matrix in a basis of $4 \times 4$
matrices, the tensor products of the identity matrix $\mathbbm{1}_2$
and the Pauli matrices $\sigma^i$,
$$
\rho= \frac{1}{4} \left( \mathbbm{1}_2 \otimes \mathbbm{1}_2 +
    a_i\,\sigma^i \otimes \mathbbm{1}_2 + b_i\,\mathbbm{1}_2 \otimes \sigma^i +
    c_{ij}\,\sigma^i \otimes \sigma^j \right)
$$
with $ a_i,b_i,c_{ij} \in \mathbbm{R}$. The parameters $a_i, b_i$
are called \textit{local} parameters as they determine the
statistics of the reduced matrices, i.e. of Alice's or Bob's system.
In order to obtain a geometrical picture as in
Ref.~\cite{BNT02,Horodecki} we consider in the following only states
where the local parameters are zero ($\vec{a}=\vec{b}=\vec{0}$),
i.e., the set of all locally maximally mixed states,
$Tr_A(\rho)=Tr_B(\rho)=\frac{1}{2}\mathbbm{1}_2$.

A state is called separable if and only if it can be written in the
form $\sum_i p_i\, \rho_i^A\otimes \rho_i^B$ with $p_i\geq 0,\sum
p_i=1$, otherwise it is entangled. As the property of separability
does not change under local unitary transformation and classical
communication (LOCC) the states under consideration can be written
in the form \cite{BNT02}
\begin{equation*}
\rho \;=\; \frac{1}{4}\left(\mathbbm{1}_2 \otimes
\mathbbm{1}_2+c_i\,\sigma^i \otimes \sigma^i \right)\, ,
\end{equation*}
where the $c_i$ are three real parameters and can be considered as a
vector  $\vec{c}$ in Euclidean space. In
Fig.~\ref{qubitstetrahedron} we show a $3$--dimensional picture,
where each point $\vec{c}$ corresponds to a locally maximally mixed
state $\rho$. The origin $\vec{c}=\vec{0}$ corresponds to the
totally mixed state, i.e. $\frac{1}{4} \mathbbm{1}_2 \otimes
\mathbbm{1}_2$. The only pure states in the picture are given by
$|\vec{c}|=3$ and represent the four maximally entangled Bell states
$|\psi^\pm\rangle=\frac{1}{\sqrt{2}}\{|01\rangle\pm|10\rangle\},|\phi^\pm\rangle=\frac{1}{\sqrt{2}}
\{|00\rangle\pm|11\rangle\}$.

It is well known that density matrices which have at least one
negative eigenvalue after partial transpose ($PT$), i.e. ${\cal
T}\otimes\mathbbm{1}$ (${\cal T}$...transposition), are entangled.
The inversion of the argument is only true for systems with
$2\otimes2$ and $2\otimes3$ degrees of freedom. $PT$ corresponds to
a reflection, i.e. $c_2\rightarrow-c_2$ with all other components
unchanged. Thus all points inside and at the surface of the
octahedron represent all separable states in the set.

\section{Bipartite Qutrits}

The description of single qutrits can be made very similar to the
one for qubits, i.e. any qutrit state $\omega\in {\cal
H}^3\equiv\mathbb{C}^3$ can then be expressed by
\begin{equation*}
\omega \;=\; \frac{1}{3} \left( \mathbbm{1}_3 + \sqrt{3} \,n_i
\,\lambda^i \right),\; n_i \in \mathbbm{R}\,, \; \sum_{i=1}^8 n_i^2
= \left| \vec{n} \right|^2 \leq 1\;,
\end{equation*} where
$\lambda^i$ ($i=1,...,8$) are the eight Gell-Mann matrices,
generalized Pauli matrices,
with properties $\mbox{Tr}\,\lambda^i = 0, \; \mbox{Tr}\,\lambda^i
\lambda^j = 2\, \delta^{ij}$. However, whereas for qubits all Bloch
vector's form a unit sphere and describe density matrices, not all
$8$--dimensional Bloch vectors for qutrits describe necessarily a
density matrix. For example $n_8=1$ and all other components equal
zero describes a matrix which is not positive definite. The full
analogy between single qubits and qutrits already fails.

For bipartite qutrits we could now try to follow an analogous way as
for qubits, i.e., consider only density matrices where the local
parameters are set to zero. However, not all locally maximally mixed
states can be decomposed into maximally entangled states, the Bell
states for qutrits. Thus we have to reduce the set of all locally
maximally mixed states further and we do that with the help of an
alternative way to generalize the Pauli matrices, i.e. by unitary
matrices which are not Hermitian, see e.g. Ref.~\cite{Weyl,01Gottesman}.\\
\section{The construction of the magical simplex $\W$}

\begin{figure}
\center{\includegraphics[width=180pt,keepaspectratio=true]{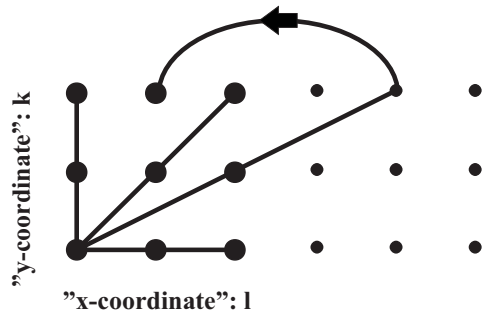}}
\caption{Here we plotted the points $P_{k,l}$ of the discrete
classical phase space. $l$ denotes the values of the position
coordinate and runs from $0$ to $2$ and $k$ ``quantizes'' the
momentum and runs also from $0$ to $2$. From one fixed point, e.g.
$P_{0,0}$, all possible lines are drawn. Thus the phase space
carries $4$ bundles where each bundle consists of $3$ parallel
lines. In Ref.~\cite{BHN1} it is shown that transformations inside
the simplex $\W$ are equivalent to transformations in this phase
space and that the lines are all equivalent in the sense that each
line may be transformed into any other one. This enables us to study
the geometry of separability and PPT in $\W$ by e.g. just
considering $3$ Bell states on a line.}\label{phasespace}
\end{figure}
We start with a maximally entangled pure state, this is a Bell type
state, in a chosen basis $\{0,1,2\}$
\begin{equation*}
\Omega _{0 0} = \frac{1}{\sqrt{3}} \sum_{s=0}^2 |s\rangle \otimes
|s\rangle\,.
\end{equation*}
On the first subspace, the system of Alice, we act with the Weyl
operators, defined by $W_{k,l}|s\rangle = w^{k(s-l)}|s-l\rangle$
with $w=e^{2\pi i/3}$, while Bob's subsystem is always left inert.
The indexes $k$ and $l$ run from $0$ to $2$. The other eight Bell
states are constructed by acting with the Weyl operators onto the
chosen Bell state
\begin{equation*} \Omega_{k,l}=W_{k,l}\otimes\mathbbm{1}_3\;
\Omega_{0,0}\,.
\end{equation*}
With that we can construct nine Bell projectors $P_{k,l}=|\Omega_{k,
l}\rangle\langle\Omega_{k,l}|$. The mixtures of these pure states
form our object of interest, the {\bf magic simplex $\W$}:
\begin{equation*} \W\;=\; \{ \quad \sum
c_{k l}\;P_{k,l} \; |\; c_{kl}\geq 0 , \quad \sum c_{k l}=1 \quad
\}\,.
\end{equation*}
Our aim is to discuss the geometry of this $8$--dimensional simplex
in the context of separability and entanglement. We focus mainly on
entanglement detected by $PT$ and give examples for a whole region
of bound entangled states, i.e. states which are $PPT$ but not
separable, for a certain class of states.

Clearly, the same construction can be used for qubits, i.e. choose
any Bell state, e.g. $\Omega_{0,0}=|\phi^+\rangle$, act on one
subspace with the Weyl operators $W_{k,l}$ where $w=e^{2\pi i/2}=-1$
and $k, l$ runs from $0$ to $1$ (equivalent to the Pauli matrices).
One obtains all four Bell states; this also generalizes for any
bipartite qudit system.

Let us remark that for qutrits not all locally maximally mixed
states can be diagonalized by Bell type states and even if so, they
may not be embedded into a version of $\W$. Moreover, there exist
nine mutually orthogonal Bell type states which do not form an
equivalent to $\W$. Examples and proofs can be found in
Ref.~\cite{BHN1,BHN2} as well as how $\W$ is embedded in the whole
state space.

Of course it is difficult to draw a picture of this $8$--dimensional
simplex, however, it can be considerably simplified because of the
high symmetry inside $\W$. This means that certain mixtures of Bell
states, $P_{k,l}$, form equivalence classes: The indexes $k$ and $l$
of the Bell states $P_{k,l}$ can be interpreted as the ``quantized''
momentum and position coordinate, respectively. In
Fig.~\ref{phasespace} we have drawn such a phase space
interpretation. It turns out that the symmetry of $\W$---appearing
as reflection, rotation and shear in the phase space---is such that
all states on a line, as indicated in Fig.~\ref{phasespace}, have
the same geometry concerning separability and entanglement. This
means states which are mixtures of the unity and three Bell states
on a line can be handled on the same footing, which is done in the
next section. Another geometrical picture is obtained if one chooses
any two Bell states, which clearly define a certain line, and any
other Bell state which does not belong to this line, this is studied
at the end.

This describes the full geometry of those states which can be
decomposed into the unity and three Bell states and is discussed in
the following. Of course to get the full geometry one has to
consider also mixtures of more than three Bell states.

\subsection{The geometry on a line}

\begin{figure}
\center{(a)\includegraphics[height=115pt,
keepaspectratio=true]{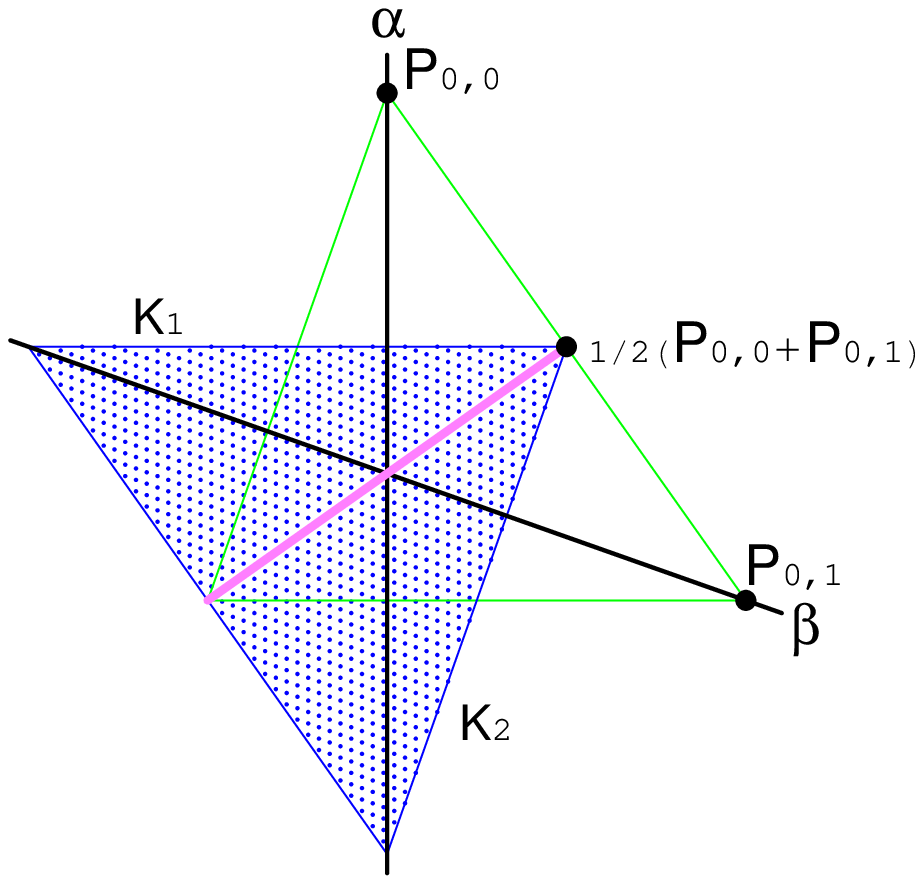}(b)
\includegraphics[height=115pt,
keepaspectratio=true]{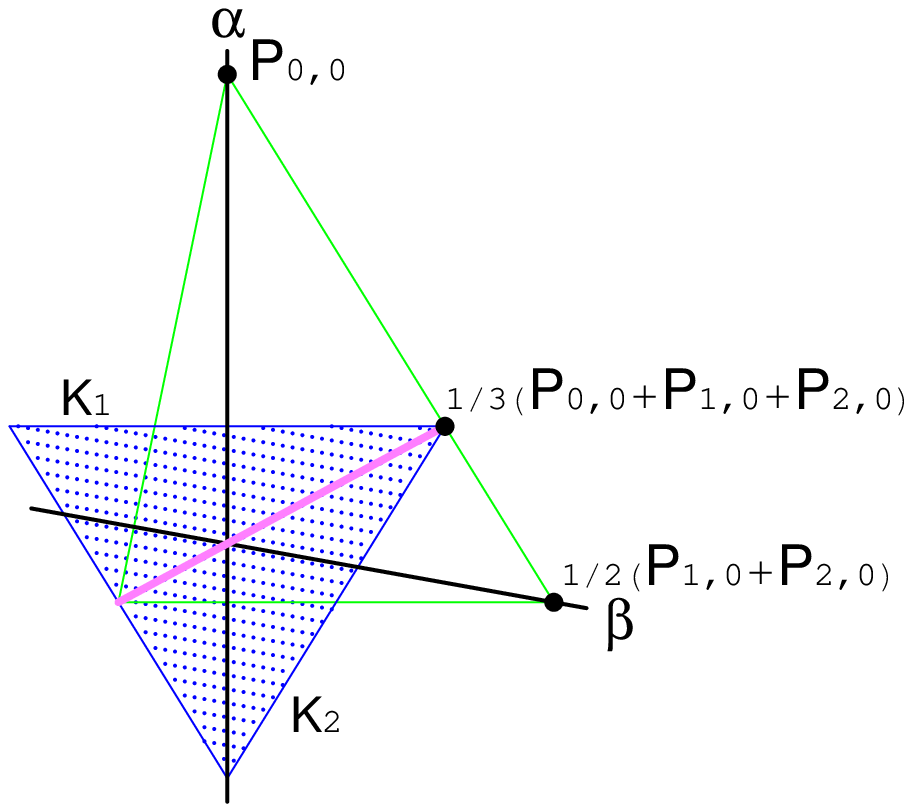}}\caption{The left figure
visualizes the geometry corresponding to the bipartite qubit states
$\rho=\frac{1-\alpha-\beta}{4}\,\mathbbm{1}_4+\alpha\,
P_{0,0}+\beta\, P_{0,1}$ and the right figure the geometry
corresponding to the bipartite qutrit states
$\rho=\frac{1-\alpha-\beta}{9}\,\mathbbm{1}_9+\alpha\,
P_{0,0}+\frac{\beta}{2}\, (P_{0,1}+P_{0,2})$. The green triangle
presents positivity and the dotted blue triangle all matrices
positive under partial transpose ($PPT$). The axes are chosen in
such a way that the symmetry of $\W$ becomes a geometrical symmetry.
In both cases it turns out that the intersection of positivity and
$PPT$ equals separability, thus the two (blue) border lines of $PPT$
crossing the axes, $K_1, K_2$, are optimal tangential witnesses for
the states represented. While for qubits this includes the whole
symmetry on a line, for qutrits we have also other symmetries, see
Fig.~\ref{secondcase}.}\label{firstcase}
\end{figure}

As a first example let us consider the states
\begin{equation}\label{example1}
\rho=\frac{1-\alpha-\beta}{9}\,\mathbbm{1}_3\otimes\mathbbm{1}_3+\alpha\,
P_{0,0}+\frac{\beta}{2}\,\left(P_{1,0}+P_{2,0}\right)\,.
\end{equation}
The geometry is given in Fig.~\ref{firstcase}~(b). The positivity
condition ($\rho \geq 0$) is satisfied for all points
$\{\beta,\alpha\}$ inside the (green) triangle (only $3$ different
eigenvalues). All states positive under $PT$, $({\cal
T}\otimes\mathbbm{1}_3)\,\rho\geq 0$, have to be inside the dotted
(blue) triangle (only $3$ different eigenvalues). The intersection
of both triangles corresponds to either separable or bound entangled
states. To find out whether the dotted area also includes bound
entangled states, one has to construct optimal tangential witnesses.
An entanglement witness for a given state $\rho$ is a criterion to
decide whether $\rho$ is inside the set of all separable states,
i.e. $\cal S$ (see also
Refs.~\cite{B05,BertlmannKrammer,BertlmannKrammer2}). The set of
tangential witnesses $K$ for a state $\rho$ is defined by
$\{K=K^\dagger\not=0|\forall\, \sigma\in {\cal S}: Tr(K
\sigma)\geq0, Tr(K\rho)=0\}$.

In the qubit case the planes of the octahedron in
Fig.~\ref{qubitstetrahedron} or the lines in
Fig.~\ref{firstcase}~(a) represent such tangential entanglement
witnesses, $K_1, K_2$, which are optimal for discriminating between
separable and entangled states. In Ref.~\cite{BHN1} it is proven
that a witness for states on a line mixed with unity must have the
form
$$K=\lambda \frac{1}{3} \mathbbm{1}_3\otimes\mathbbm{1}_3+\sum_k
\kappa_k P_{k,0}\,.$$ Furthermore, $K$ is directly related to the
matrices $$M_\Phi=\lambda \mathbbm{1}_3+\sum_k \kappa_k
W_{k,0}|\Phi\rangle\langle\Phi|W_{k,0}^\dagger$$ with $\Phi$ being
any normalized state vector in $\mathbb{C}^3$. If $ M_\Phi$ is
non--negative $\forall\;\Phi$, $K$ is a witness and moreover if
$\det M_\Phi=0$, then $K$ is a tangential witness.

For the states~(\ref{example1}) it turns out that $K_1, K_2$---the
limiting lines of $PT$--- are optimal tangential witnesses, i.e.
$\det M_\Phi=0$ for certain $\Phi$'s. This means the dotted area
inside positivity represents separable states, i.e.
$PPT\equiv$separable, see Fig.~\ref{firstcase}~(b).

\begin{figure}
\center{\includegraphics[height=180pt,
keepaspectratio=true]{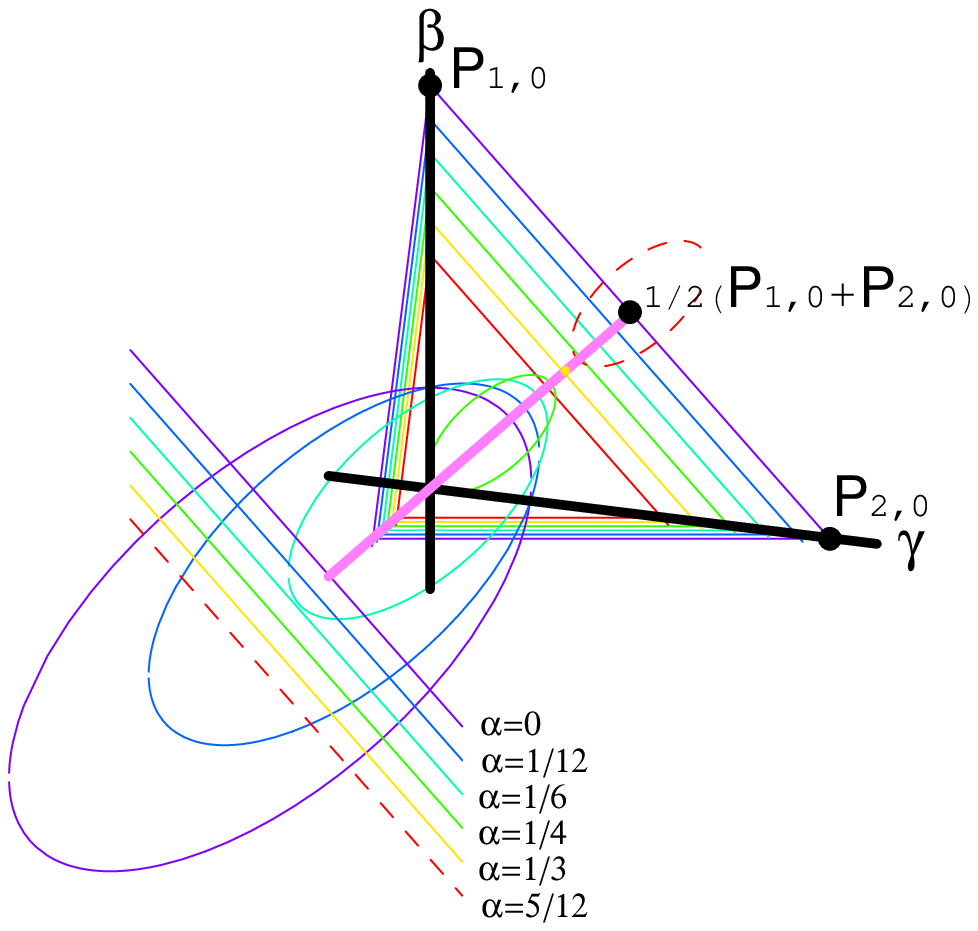}\includegraphics[height=140pt,
keepaspectratio=true]{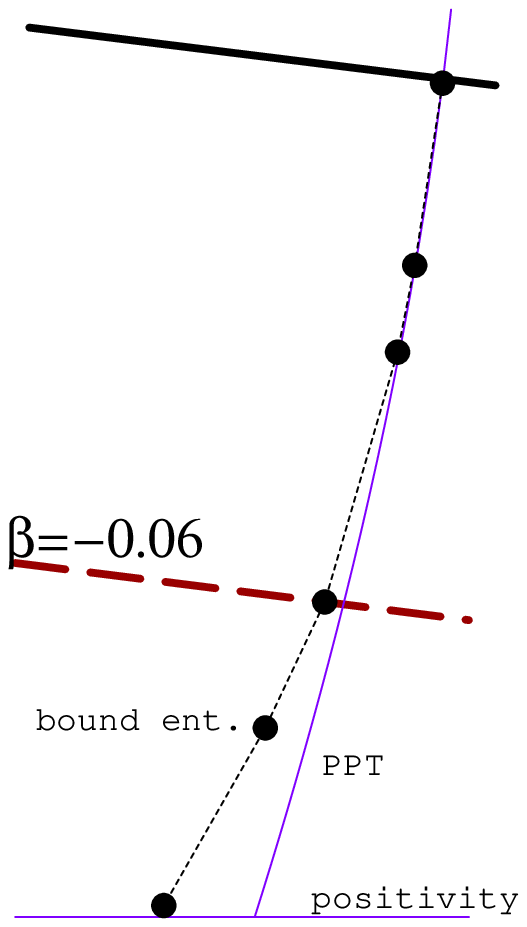}} \caption{The left figure
visualizes the slices through the state space
$\rho=\frac{1-\alpha-\beta-\gamma}{9}\,\mathbbm{1}_3\otimes\mathbbm{1}_3+\alpha
P_{0,0}+\beta\, P_{1,0}+\gamma P_{2,0}$, where the biggest triangle
($\equiv$ positivity) and the biggest ellipse/line ($\equiv$ $PPT$)
correspond to $\alpha=0$ (purple), the next biggest objects to
$\alpha=\frac{1}{12}$ (blue) and so on until $\alpha=\frac{5}{12}$
(dashed, red). For $\alpha<\frac{1}{6}$ the $PPT$ region is an
ellipse cut by a line and for $\frac{1}{6}\leq\alpha<\frac{5}{12}$
it is solely an ellipse. In the last case ($\alpha=\frac{5}{12}$)
the $PPT$ area does not intersect the positivity area (smallest, red
triangle), that means all states are entangled, in agreement to
Fig~\ref{firstcase}~(b). Contrary to the slice in
Fig~\ref{firstcase}~(b) not all $PPT$ states are separable. In the
figure to the right hand side an enlargement for negative $\beta$
and $\alpha=0$ is shown. The dots are the density matrices for which
the tangential witness is optimized. That means all states between
the iterated curve and the $PPT$ ellipse correspond to bound
entangled states.}\label{secondcase}
\end{figure}

%
%
Generally, any state on a line and unity is given by
\begin{equation}
\rho=\frac{1-\alpha-\beta-\gamma}{9}\,\mathbbm{1}_3\otimes\mathbbm{1}_3+\alpha\,
P_{0,0}+\beta\, P_{1,0}+\gamma\, P_{2,0}\,.
\end{equation}
and the geometry is drawn in Fig.~\ref{secondcase}. The positivity
condition on the three eigenvalues forms again a triangle, the
condition on $PT$, however, forms a more complicated object (an
ellipse and a line). As one mixes more and more $P_{0,0}$ to the
state, the region of positivity and $PPT$ decreases, until for
$\alpha>\frac{1}{3}$ both regions do not intersect anymore, i.e.
only entangled states are found. For $\alpha=\frac{1}{3}$ a single
point, the line state, is separable. \textbf{Again we have to ask
whether the states positive under $PT$ are all separable?}

In Ref.~\cite{BHN1} the case $\alpha=0$ is discussed and indeed it
turns out that there is a small region of bound entangled states if
either $\beta$ or $\gamma$ is negative (see enlarged region, right
hand side of Fig.~\ref{secondcase}). The difference between the
$PPT$ boundary and separability is rather small, at most of the
order $10^{-2}$. \textbf{The question arises whether the bound
entangled region increases or decreases when $P_{0,0}$ is more and
more mixed to the state.}

\begin{figure}
\center{\includegraphics[height=120pt,
keepaspectratio=true]{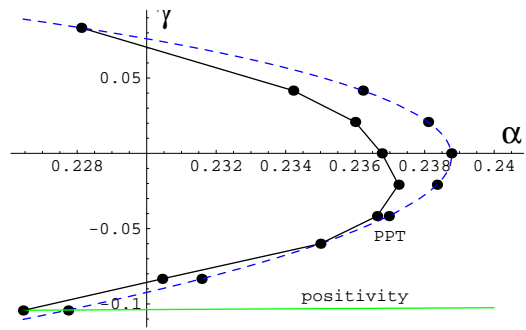}} \caption{The figure visualizes
the slices through the state space
$\rho=\frac{1-\alpha-\beta-\gamma}{9}\,\mathbbm{1}_3\otimes\mathbbm{1}_3+\alpha\,
P_{0,0}+\beta\, P_{1,0}+\gamma\, P_{2,0}$ with $\beta=-0.06$. On the
horizontal axis $\alpha$ and on the vertical axis $\gamma$ is
plotted. The (green) line represents the positivity border and the
dashed  (blue) curve the $PPT$ border. The points are derived by
optimizing the witness and therefore represent the separability
border. Between the two curves one has the bound entangled states.
Note that clearly for $\beta=\gamma=-0.06$ the state is not bound
entangled for any $\alpha$, see
Fig.~\ref{firstcase}~(b).}\label{boundforbeta}
\end{figure}

Here we can distinguish two cases, i.e. $P_{0,0}$ is mixed with
positive or negative $\alpha$. If we choose for instance
$\beta=-0.06$ (see also Fig.~\ref{boundforbeta}) we find that for
$\alpha\geq 0$ the region decreases, i.e. already for $\alpha=1/12$
we find no better witness than given by $PPT$ up to numerical
precision of $10^{-6}$ (using standard optimization methods). For
$\alpha\leq 0$ the region decreases until for $\beta=\gamma=-0.06$
no bound entangled state is found and then increases again. Clearly
in the case $\beta=\gamma=-0.06$ we have the case represented by
Fig.~\ref{firstcase}~(b) where no bound entangled states can be
found.


The ``generalized'' concurrence for qutrits \cite{Buchleitner} turns
out to give the same result for $\gamma=0$ for certain numerical
quasi pure approximations \cite{BuchleitnerSauer}. Also for
$\gamma=\pm\frac{1}{12}$ the results coincide.

Summarizing, the geometry of mixtures of three Bell states on a line
and the unity is such that separable and bound entangled states form
sections through the simplex, where the bound entangled states form
only a small region, see also Fig.~\ref{firstcase}~(b),
Fig.~\ref{secondcase} and Fig.~\ref{boundforbeta}.


\subsection{The geometry beyond lines}

\begin{figure}
\center{(a)\includegraphics[height=110pt,
keepaspectratio=true]{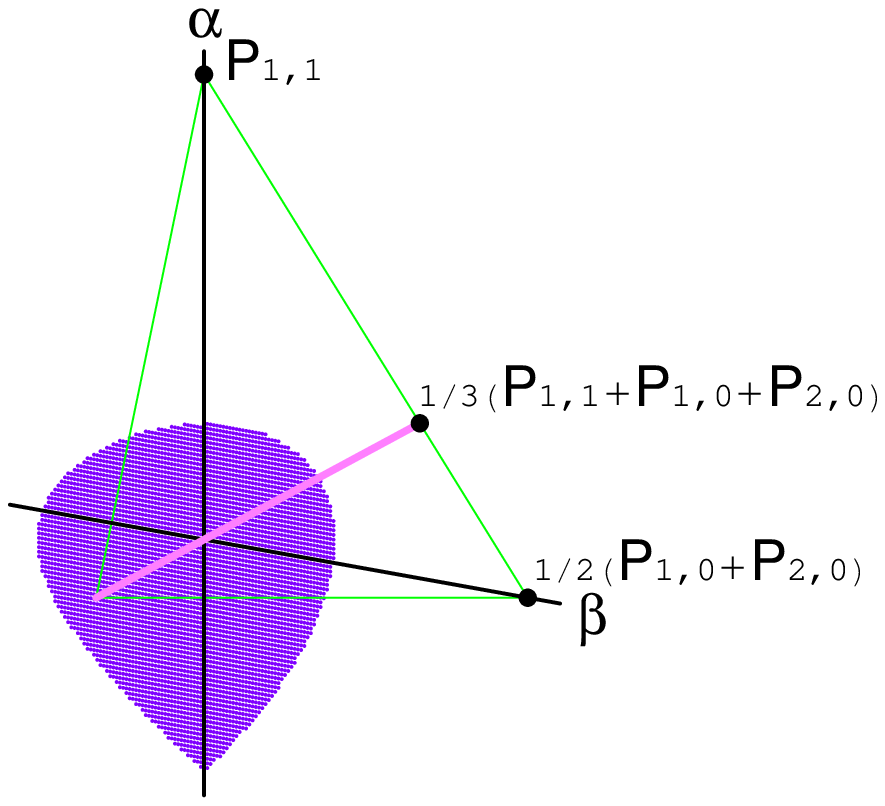}
(b)\includegraphics[height=110pt,
keepaspectratio=true]{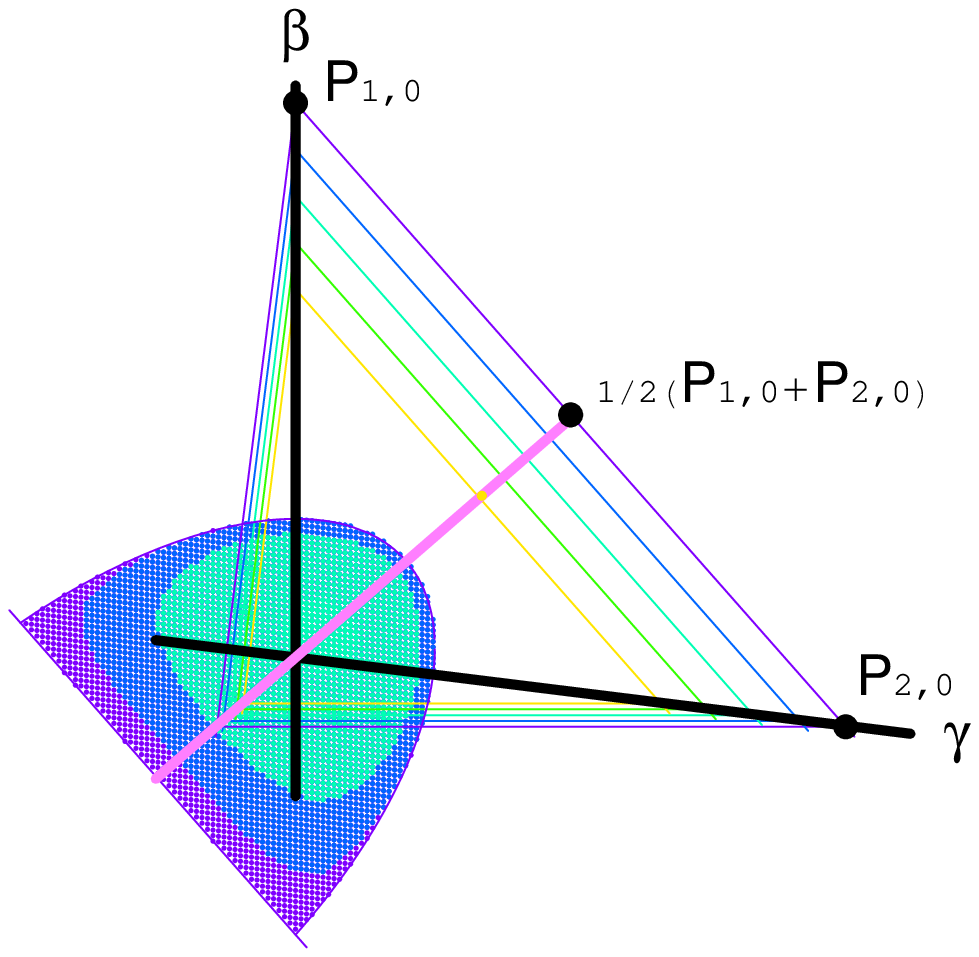}}\caption{Both figures
visualize the state space
$\rho=\frac{1-\alpha-\beta-\gamma}{9}\,\mathbbm{1}_9+\alpha\,
P_{10}+\beta\, P_{20}+\gamma\, P_{11}$, where in (a)
$\beta,\gamma\rightarrow \frac{\beta}{2}$ and in (b) slices with
$\alpha=0,\frac{1}{12},\frac{1}{6},\frac{1}{4},\frac{1}{3}$ are
shown (color coding as in Fig.~\ref{secondcase}). Compared to the
line case, Fig.~\ref{firstcase}~(b) and Fig.~\ref{secondcase}, the
$PPT$ region is more complicated, i.e. is not simply an ellipse. The
$PPT$ region shrinks with increasing $\alpha$ and already for
$\alpha=\frac{1}{3}$ (smallest (yellow) triangle) no separable state
is found.
}\label{thirdcase}
\end{figure}

The second possibility of a mixture of $3$ Bell states is to choose
two Bell states which always define a line and choose any other Bell
state which is not the one completing the line and mix it with the
unity, e.g.
\begin{equation}
\rho=\frac{1-\alpha-\beta-\gamma}{9}\,\mathbbm{1}_3\otimes\mathbbm{1}_3+\alpha\,
P_{10}+\beta P_{20}+\gamma P_{11}\,.
\end{equation}
Clearly, the positivity condition gives the same three different
eigenvalues as in the cases before. However, if one chooses
$\beta,\gamma\rightarrow \frac{\beta}{2}$, the boundary of $PPT$
consists no longer of simple lines, see Fig.~\ref{thirdcase}~(a). If
one considers slices spanned up by the unity and two Bell states,
Fig.~\ref{thirdcase}~(b), we obtain no longer a cone but a more
complex object.

In principle any optimal witness can be calculated with the
procedure as given in Ref.~\cite{BHN1}, more parameters, however,
are involved and consequently it is hard to minimize $\det M_\Phi$.
For that another numerical strategy is necessary and will be done in
a future work. Calculation with the `generalized'' concurrence for
qutrits \cite{Buchleitner} for certain numerical quasi pure
approximations \cite{BuchleitnerSauer} have shown that there are
also bound entangled regions in this case.

\section{Summary}

We discuss the state space of locally maximally mixed bipartite
qubits, i.e. such states where the trace over one party gives the
normalized unity. A three dimensional picture of the geometry of
separability and entanglement can be drawn,
Fig.~\ref{qubitstetrahedron}. We generalize for bipartite qutrits,
where a certain smaller state space, a subspace of all locally
maximally mixed density matrices, is considered. We obtain it by
acting with the Weyl operators, generalized Pauli matrices, onto a
chosen maximally entangled state, a Bell state. In this way one can
construct nine Bell states and its convex combination forms our
object of interest, the magic simplex $\W$.


It has nine Bell states in the corners and given the high symmetry
of this state space, one finds a phase--space structure, see
Fig.~\ref{phasespace}, implying that certain Bell states form
equivalency classes.

We investigate the geometry of $\W$, in particular the geometry of
three such Bell states forming a line mixed with the unity, see
Fig.~\ref{firstcase} and Fig.~\ref{secondcase}. We find that the
separable states do not form a simple polytope as in the qubit case
and despite the fact that only a subset of all locally maximally
mixed states is considered we find even a whole region of bound
entangled states. They are obtained by optimizing entanglement
witnesses. The high symmetry on a line reduces the class of
witnesses remarkably and consequently optimization is numerically
obtainable. Furthermore, we find that the region of bound entangled
states decreases/increases if the third state of the line is mixed
to it.

Last but not least we investigate the state space of two Bell
states, the unity and another Bell state not on the line formed by
the previous ones. The region representing the states positive under
partial transpose ($PPT$) show a even more complicated geometry, see
Fig.~\ref{thirdcase}.

Concluding, while for bipartite qubits the geometry of separability
and entanglement is exhausted by considering a line --the mixture of
the unity and any two Bell states-- for bipartite qutrits the
geometry is more complicated and includes new phenomena which may
find interesting applications in future.

\textbf{Acknowledgements:} Many thanks to Simeon Sauer, Fernando de
Melo, Joonwoo Bae, Florian Mintert and Andreas Buchleitner for
providing us with their numerical results of the ``generalized''
concurrence.


\begin{thebibliography}{1}

\bibitem{Bennett}
C.H. Bennett, G. Brassard, S. Popescu, B. Schumacher, J. Smolin, and
W.K. Wootters, Phys. Rev. Lett. {\bf 78}, 2031 (1996).

\bibitem{N06} H. Narnhofer
, J. Phys. A {\bf 39}, 7051 -- 64.

\bibitem{VW00} K. G. H. Vollbrecht, R. F. Werner
, Phys. Rev. A {\bf 64}, 062307 (2001).

\bibitem{Ovrum} J.M. Leinaas, J. Myrheim, E. Ovrum, Phys. Rev. A {\bf 74}, 012313
(2006).

\bibitem{PR04} O. Pittenger, M.H. Rubin
, Lin. Alg. Appl. {\bf 390}, 255 (2004).


\bibitem{W04} W.K. Wootters:
{\it Quantum measurements and finite geometry};
arXiv:quant-ph/0406032.

\bibitem{B04} I. Bengtsson:
{\it MUBs, polytopes and finite geometries}; arXiv:quant-ph/0406174.

\bibitem{VW99} K.G.H. Vollbrecht, R.F. Werner:
J. Math Phys. {\bf 41}, 6772 (2000).


\bibitem{Horodecki1998}
M. Horodecki, M. Horodecki, and R. Horodecki, Phys. Rev. Lett. {\bf
80}, 5239 (1998).


\bibitem{Augusiak}
R. Augusiak, J. Stasin´ska and P. Horodecki, ``{\it Beyond the
standard entropic inequalities: stronger scalar separability
criteria and their applications}'', arXiv: 0707.431.

\bibitem{Breuer}
H.-P. Breuer, Phys. Rev. Lett. {\bf 97}, 080501 (2006).

\bibitem{Derkacz}
L. Derkacz and L. Jakobczyk. ``{\it Entanglement of a class of mixed
two--qutrit states}'', arXiv: 0707.1575.

\bibitem{Ruskai}
M. Nathanson and M.B: Ruskai, Mathematical Systems Theory {\bf 40},
8171 (2007).

\bibitem{Planat}
M.R.P. Planat, H. Rosu, S. Perrine and M. Saniga, Foundations of
Physics {\bf 36}, 1662 (2006).

\bibitem{Acin}
D. Cavalcanti, A. Ferraro, A. Garcia-Saez and A. Acin, {\it Thermal
bound entanglement and area laws}, arXiv: 0705.3762.

\bibitem{HorodeckiSummary}
R. Horodecki, P. Horodecki, M. Horodecki and K. Horodecki, ``{\it
Quantum entanglement}'', arXiv: 0702.225.

\bibitem{Horodecki}
R. Horodecki and M. Horodecki, Phys. Rev. A {\bf 54}, 1838 (1996).

\bibitem{BNT02} R.A. Bertlmann, H. Narnhofer, and W. Thirring,
Phys. Rev. A {\bf 66}, 032319 (2002).


\bibitem{Weyl} H. Weyl:
{\it Gruppentheorie und Quantenmechanik}, zweite Auflage, (S.
Hirzel, Leipzig, 1931).

\bibitem{01Gottesman}
D. Gottesman, A. Kitaev, and J. Preskill, 
Phys. Rev. A \textbf{64}, 012310 (2001).

\bibitem{BHN1}
B. Baumgartner, B.C. Hiesmayr, and H. Narnhofer, Phys. Rev. A {\bf
74}, 032327 (2006).

\bibitem{BHN2}
B. Baumgartner, B.C. Hiesmayr, and H. Narnhofer, 
J. Phys. A {\bf 40}, 7919 (2007).

\bibitem{B05} R.A. Bertlmann, K. Durstberger, B.C. Hiesmayr and P.
Krammer,
Phys. Rev. A {\bf 72}, 052331 (2005).

\bibitem{BertlmannKrammer}
R.A. Bertlmann and Ph. Krammer, ``{\it Bloch vectors for qudits and
geometry of entanglement}'', arXiv: 0706.1743.

\bibitem{BertlmannKrammer2}
R.A. Bertlmann and Ph. Krammer, ``{\it Geometric entanglement
witnesses and bound entanglement}'', arXiv: 0710.1184.

\bibitem{Buchleitner}
F. Minert, M. Kus and A. Buchleitner, 
Phys. Rev. Lett. {\bf 92}, 167902 (2004).

\bibitem{BuchleitnerSauer}
S. Sauer, F. de Melo, J. Bae, F. Mintert and A. Buchleitner, {\it
private communication}.


%
%
%

%
%

%

%
%
%
%
%
%
%






\end{thebibliography}
\end{document}